\documentclass[aps,longbibliography,nofootinbib,showkeys,12pt]{revtex4-2}

\usepackage{amssymb}
\usepackage{color}

\usepackage{fullpage}

\def\thetacirc{\stackrel{\,\,\circ}{\theta}}

\def\psic{\stackrel{\,\,\circ}{\psi}}

\begin{document}

\title{Coherent states for equally spaced, homogeneous waveguide arrays}
\author{Julio Guerrero}
\email{jguerrer@ujaen.es: corresponding author}
\affiliation{Department of Mathematics, University of Ja\'en, Campus Las Lagunillas s/n, 23071 Ja\'en, Spain}
\affiliation{Institute Carlos I of Theoretical and Computational Physics (iC1), University of  Granada,
Fuentenueva s/n, 18071 Granada, Spain}

\author{H\'ector M. Moya-Cessa}
\email{hmmc@inaoep.mx}
\affiliation{Instituto Nacional de Astrof\'{\i}sica, \'Optica y Electr\'onica, INAOE, Puebla, Mexico}

\date{\today}

\begin{abstract}
\section*{Abstract}
Coherent states for equally spaced, homogeneous waveguide arrays are defined, in the infinite, semiinfinite and finite cases, and resolutions of the identity are constructed, using different methods. In the infinite case, which corresponds to Euclidean coherent states, 
a resolution of the identity with coherent states on the circle and involving a nonlocal inner product is reviewed. In the semiinfinite case, which corresponds to London coherent states, various 
construction are given (restricting to the circle with a non-local scalar product, rescaling the coherent states, modifying them, or using a non-tight frame). In the finite case, a construction in terms of coherent states on the circle is given, and this construction is shown to be  a regularization of the infinite and semiinfinite cases.
\end{abstract}

\maketitle

\section{Introduction}

Many efforts have been directed towards the engineering of nonclassical states of quantum mechanical systems, in which certain observables exhibit less fluctuations than that of a coherent state, the so-called standard quantum limit (SQL).  In this context,  coherent states for the electromagnetic field, introduced by Glauber \cite{Glauber63a,Glauber63b} and Sudarshan \cite{Sudarshan63}, are of importance because  their relatively easy way of generation in the laboratory and their classical-like wave behaviour. These states may be obtained by
different mathematical definitions: a) As eigenstates of the annihilation operator b)  by the application of the displacement operator on the vacuum state of the harmonic oscillator, c) as states for which their time-evolving wave function shape does not change in time and whose centroid follows the motion of a classical point particle in a harmonic oscillator potential and d)  coherent states that have a resolution to the identity \cite{Klauder63-I,Klauder94}.

By using harmonic oscillator algebras, one may obtain the same coherent states from the three definitions above. However, for more complicated dynamical systems, beyond the harmonic oscillator, there is a need for  new methods to generalize the idea of coherent states. Nieto and Simmons \cite{Nieto78-I,Nieto79-II,Nieto79-III} have constructed coherent states for potentials whose energy spectra have unequally spaced energy levels, such as the Morse potential, the Poschl-Teller potential and the harmonic oscillator with centripetal barrier.

Man’ko {\it et al.} \cite{Manko97} introduced coherent states of an $f$-deformed algebra as eigenstates of a generalized annihilation operator, {\it i.e.,} an annihilation operator composed by the annihilation operator for the harmonic oscillator times a function of the number operator. A remarkable result is that such  states may present nonclassical properties as squeezing and antibunching \cite{Vogel96}.

Coherent states may be modelled in classical systems \cite{LeonMontiel11a,LenMontiel11b} by injecting classical light in equally spaced waveguide arrays and monitoring its propagation properties. In particular it has been shown that Susskind-Glogower (or London) coherent states may be produced in semi-infinite arrays of waveguides when light is injected in the first waveguide (that corresponds to the vacuum state) \cite{LeonMontiel11a,LenMontiel11b} (see also 
\cite{Curado21,Zelaya21}). In the case of an infinite array (also equally spaced) coherent states of the group E(2) may be generated, whose amplitudes are given by Bessel functions that depend on the propagation distance \cite{Jones65} (see also \cite{Hacyan08} for other coherent states containing Bessel functions). On the other hand, a finite system of equally spaced waveguide arrays may be solved analytically with the use of Chebyshev polynomials \cite{Makris06,SotoEguibar11}.

In this paper we shall focus on coherent states in equally spaced, homogeneous waveguide arrays, either infinite, semi-infinite and finite, identifying the obstacles in their construction and proposing different possibilities for a resolution of the identity. In particular, it will be shown that in all cases it is possible to construct a resolution of the identity with coherent states restricted to the circle, as it is has been already done for the infinite case \cite{Guerrero21}.  In fact, both the infinite and semiinfinite cases are easily recovered from the finite case when the number of waveguides grows to infinite, in this sense the finite case can be used as a regularization procedure for the infinite and semiinfinite cases.

\section{Equally spaced, homogeneous waveguide arrays}
\label{WGA}

As commented in the introduction, waveguide arrays are a good testbed to simulate both classical and quantum phenomena \cite{Mar-Sarao08,PerezLeija10,PerezLeija12, RodriguezLara14,RodriguezLara15}. We shall discuss in this section, to motivate physically the coherent states discussed in this paper, the model of equally spaced and homogeneous waveguide arrays, in the cases of an infinite, a semi-infinite and a finite number of waveguides.

\subsection{Infinite case}
\label{WGA-infinite}


Consider first an infinite waveguide array of equally spaced and homogeneous optical single mode waveguides. The equations describing  light propagation along the z-direction are given by (see \cite{Jones65}):
\begin{equation}
 i\frac{d A_n}{d z} = A_{n+1}+A_{n-1}\,,\qquad n\in\mathbb{Z}\,,
 \label{CoupledEqinf}
\end{equation}
where $A_n(z)$ is the electric modal field in the $n$-th waveguide at position $z$. The distance $z$ is measured in units of the coupling constant between waveguides (supposed constant) and the propagation constant has been removed by a suitable transformation (see \cite{Makris06}).

To describe algebraically this situation, let us use Dirac notation (see \cite{RodriguezLara14}) and introduce an abstract Hilbert space ${\cal H}$, expanded by the extended Fock basis $\bar{\cal F}=\{|n\rangle,\,n\in\mathbb{Z}\}$, where the Fock state $|n\rangle$ represents the $n$-th waveguide, and that is isomorphic to $\ell^2(\mathbb{Z})$.

Therefore the infinite length vector  ${\cal A}=(\ldots,A_0,\ldots)$ can be represented (in Dirac's notation)  as:
\begin{equation}
  |{\cal A}\rangle=\sum_{n\in\mathbb{Z}} A_n|n\rangle\,,
\end{equation}

 Elements in ${\cal H}$ represent amplitude of light distributions along the whole waveguides with finite total energy for each value of $z$.
\begin{equation}
 ||{\cal A}(z)||^2 = |||{\cal A}(z)\rangle||^2=\sum_{n\in\mathbb{Z}} |A_n(z)|^2 < \infty\,.
\end{equation}

The Eqs. (\ref{CoupledEqinf}) can be written as a Schr\"odinger equation, with $-z$ playing the role of time:
\begin{equation}
 -i \frac{d\ }{d z}|{\cal A}\rangle = \hat{H}  |{\cal A}\rangle\,,
\end{equation}
with the Hamiltonian given by $\hat{H} =\hat{V}^\dagger+ \hat{V}$,
where $\hat{V}^\dagger$ and $\hat{V}$ are step operators:
\begin{equation}
\hat{V}|n\rangle = |n-1\rangle\,,\qquad
 \hat{V}^\dagger|n\rangle = |n+1\rangle\,,\qquad n\in\mathbb{Z}\,.
\end{equation}

The propagator $\hat{U}(z)$, verifying $|{\cal A}(z)\rangle = \hat U(z)|{\cal A}(0)\rangle$, turns out to be \cite{Jones65}:
\begin{equation}
 \hat{U}(z)=e^{i z \hat{H}}=\sum_{n,m\in \mathbb{Z}} i^{n-m} J_{n-m}(2z)|n\rangle\langle m|\,.
\end{equation}

It is interesting to note that if we define the \textit{twisted} Hamiltonian\footnote{This kind of Hamiltonian describes  waveguides that are twisted around each other \cite{TwistedWG15}, although this is only physically realizable in the case of a finite number of waveguides. For convenience, the twisting has been shifted by $\frac{\pi}{2}$, so the non twisted case is recovered for $\theta=\frac{\pi}{2}$.}:
\begin{equation}
\hat{H}^\theta = e^{i (\theta-\frac{\pi}{2} )\hat{n}}\hat{H}e^{-i (\theta -\frac{\pi}{2})\hat{n}}=  
-i (e^{i\theta} \hat{V}^\dagger-e^{-i\theta}\hat{V})\,, \label{Htwisted}
\end{equation}
where $\hat{n}|n\rangle = n|n\rangle\,,\,n\in\mathbb{Z}$, is the number operator, then the corresponding \textit{twisted} propagator is
\begin{equation}
 \hat{U}^\theta(z)=e^{i z \hat{H}^\theta} = e^{z( e^{i\theta} \hat{V}^\dagger-e^{-i\theta}\hat{V}   )}=\sum_{n,m\in \mathbb{Z}} e^{i(n-m)\theta} J_{n-m}(2z)|n\rangle\langle m|\,.
\end{equation}

In particular, the original Hamiltonian $\hat H$ and propagator $\hat U$ are recovered for $\theta=\frac{\pi}{2}$. Also note that we can restrict the twisting $\theta$ to $[0,\pi)$, since the case $\theta\in[\pi,2\pi)$ is obtained by backward propagation with $\theta-\pi$.

In Sec. \ref{E(2)-CS} we discuss  E(2) coherent states, which are generated by a displacement  operator that coincides with  $\hat{U}^\theta$.

\subsection{Semi-infinite case}
\label{WGA-semiinfinite}

In the case of a semi-infinite number of equally spaced, homogeneous waveguides, the equations describing light propagation are given by (see, for instance, \cite{Makris06}):
\begin{equation}
 i\frac{d A_n}{d z} = A_{n+1}+A_{n-1}\,,\quad n\geq 1\,,\qquad i\frac{d A_0}{d z} = A_{1} \,.
 \label{CoupledEqsemiinf}
\end{equation}

In this case consider the Hilbert subspace ${\cal H}_+$, expanded by the usual Fock basis
${\cal F}=\{|n\rangle,\,n\geq 0\}$. Consider the orthogonal projector $\hat{P}_+$ onto this subspace, and define the truncated step operators $\hat V_+= \hat P_+ \hat V$ and $\hat V^\dagger_+= \hat V^\dagger \hat P_+ $, satisfying
\begin{eqnarray}
\hat{V}_+|n\rangle & =& |n-1\rangle\,,\quad n\geq 1\,,\qquad \hat{V}_+|0\rangle = 0\,,\nonumber\\
 \hat{V}^\dagger_+|n\rangle &=& |n+1\rangle\,,\qquad n\geq 0\,.
\end{eqnarray}

Then the Hamiltonian in this case can be written as $\hat{H}_+ =\hat{V}^\dagger_++ \hat{V}_+$. The propagator in this case is $\hat U_+(z)=e^{i z \hat H_+}$, and its expression can be obtained by the method of the images\footnote{That is, an infinite array is considered and for each source of light at the waveguide $m$ at $z=0$, an ``image'' source, with negative amplitude, is considered at the waveguide $-m-2$, i.e. the light distribution is antisymmetric with respect to the waveguide $n=-1$ for $z=0$, $A_{-n-2}(0)=-A_n(0)$, and this property is preserved for all values of $z$, allowing to restrict the solution to $n\geq 0$.}  \cite{Makris06}:
\begin{equation}
 \hat U_+(z) = \sum_{n,m=0}^\infty \left( i^{n-m} J_{n-m}(2z)+ i^{n+m} J_{n+m+2}(2z) \right)|n\rangle\langle m|\,.
\end{equation}

The twisted versions of the Hamiltonian and propagator\footnote{In this case the method of the images can also be applied, but the resulting light distribution must have a twisted symmetry with respect to the waveguide $n=-1$, namely $A_{-n-2}(0)=-e^{-2i(\theta-\frac{\pi}{2})(n+1)} A_n(0)$, and this property is preserved under propagation in $z$, allowing to restrict the solution to $n\geq 0$.} are given by:
\begin{eqnarray}
 \hat H_+^\theta &=& e^{i (\theta-\frac{\pi}{2} )\hat{n}}\hat{H}_+e^{-i (\theta -\frac{\pi}{2})\hat{n}}=  
-i (e^{i\theta} \hat{V}^\dagger_+-e^{-i\theta}\hat{V}_+)\label{Ham-semiinf-twisted} \\
\hat{U}^\theta_+(z)&=&e^{i z \hat{H}^\theta_+} = e^{z( e^{i\theta} \hat{V}^\dagger_+-e^{-i\theta}\hat{V}_+   )}=\sum_{n,m=0}^\infty  e^{i(n-m)\theta}  \left(J_{n-m}(2z)+(-1)^m J_{n+m+2}(2z) \right)|n\rangle\langle m|\,. \label{Prop-semiinf-twisted}
\end{eqnarray}

\subsection{Finite case}
\label{WGA-finite}

Finally, in the case of a finite number $N$ of equally spaced, homogeneous waveguides, the equations describing light propagation are given by \cite{Makris06}:
\begin{eqnarray}
 i\frac{d A_n}{d z} &=& A_{n+1}+A_{n-1}\,,\qquad 1\leq n\leq N-1 \nonumber\\
 i\frac{d A_0}{d z} &=& A_{1} \\
 i\frac{d A_{N-1}}{d z} &=& A_{N-2} \nonumber
 \label{CoupledEqfinite}
\end{eqnarray}

In this case consider the finite-dimensional Hilbert subspace ${\cal H}_N$, expanded by the truncated  Fock basis
${\cal F}_N=\{|n\rangle,\,0\leq n\leq N-1\}$. Consider the orthogonal projector $\hat{P}_N$ onto this subspace, and define the truncated step operators $\hat V_N= \hat P_N \hat E\hat P_N$ and $\hat V_N^\dagger= \hat P_N\hat E^\dagger \hat P_N $, satisfying
\begin{eqnarray}
\hat{V}_N|n\rangle & =& |n-1\rangle\,,\quad 1\leq n\leq N-1\,,\qquad \hat{V}_N|0\rangle = 0\,,\nonumber\\
 \hat{V}_N^\dagger|n\rangle &=& |n+1\rangle\,,\quad 0\leq n\leq N-2\,.,\qquad \hat{V}_N^\dagger|N-1\rangle = 0\,.
\end{eqnarray}

Then the Hamiltonian in this case can be written as $\hat{H}_N =\hat{V}_N^\dagger+ \hat{V}_N$. The propagator in this case is $\hat U_N(z)=e^{i z \hat H_N}$, and its expression can be obtained again by the method of the images\footnote{That is, an finite array is considered and for each source of light at the waveguide $m$, $0\leq m\leq N-1$, at $z=0$, an infinite number of ``image'' sources, obtained by the reflection with respect to the guides $n=-1$ and $n=N$, i.e.
$A_{-n-2+(2N+2)r}(0)=-A_n(0)$ and $A_{n+(2N+2)r}(0)=A_n(0)$, for $r\in\mathbb{Z}$, in such a way
that  light intensity is zero at the waveguides $n=-1$ and $n=N$ for $z=0$, and this property is preserved for all values of $z$, allowing to restrict the solution to $0\leq n\leq N-1$.}  \cite{Makris06} or computing the eigenmodes in terms of Chebyshev polynomials \cite{SotoEguibar11}:
\begin{eqnarray}
 \hat U_N(z) &=& \sum_{n,m=0}^{N-1} \sum_{l=-\infty}^\infty i^{-(2N+2)l} \left( i^{n-m} J_{n-m-(2N+2)l}(2z)- i^{n+m+2} J_{n+m+2-(2N+2)l}(2z) \right)|n\rangle\langle m| \nonumber \\
 &=& \sum_{n,m=0}^{N-1} \sum_{\alpha=0}^{N-1} \frac{ U_{n+1}\left[\cos\left( \frac{\alpha \pi}{N+1}  \right)\right] U_{m}\left[\cos\left( \frac{\alpha \pi}{N+1}  \right)\right] exp\left[2i z \cos\left( \frac{\alpha \pi}{N+1}  \right)\right]}{\sum_{k=0}^{N-1}U_k\left[\cos\left( \frac{\alpha \pi}{N+1}  \right)\right]^2}|n\rangle\langle m|
\end{eqnarray}

The twisted versions of the Hamiltonian and propagator\footnote{In this case the method of the images can also be applied, but the resulting light distribution must have a twisted symmetry with respect to the waveguide $n=-1$ and $n=N$, namely $A_{-(n+2-2(N+1)r)}(0)=-e^{2i(\theta-\frac{\pi}{2})((n+1-(N+1)r))} A_n(0)$ and $A_{n+2(N+1)r)}(0)=e^{-2i(\theta-\frac{\pi}{2})((N+1)r))} A_n(0)\,,\forall r\in\mathbb{Z}$, with  this property being preserved under propagation in $z$, allowing to restrict the solution to $0\leq n\leq N-1$.} are given by:
\begin{eqnarray}
 \hat H_N^\theta &=& e^{i (\theta-\frac{\pi}{2} )\hat{n}}\hat{H}_Ne^{-i (\theta -\frac{\pi}{2})\hat{n}}=  
-i (e^{i\theta} \hat{V}_N^\dagger-e^{-i\theta}\hat{V}_N)\nonumber \\
\hat{U}_N^\theta(z)&=&e^{i z \hat{H}_N^\theta} =e^{z( e^{i\theta} \hat{V}^\dagger-e^{-i\theta}\hat{V}   )} \\
&=& \sum_{n,m=0}^{N-1} \sum_{l=-\infty}^\infty i^{-(2N+2)l} e^{i(n-m)\theta} \left( J_{n-m-(2N+2)l}(2z)+(-1)^{m} J_{n+m+2-(2N+2)l}(2z) \right)|n\rangle\langle m|\,. \nonumber
\end{eqnarray}

\section{E(2) coherent states}
\label{E(2)-CS}


Coherent states for the case of an infinite-dimensional array of waveguides have been constructed in \cite{Guerrero21}. Here we shall resume the main results in order to fix notation an to better compare with the semiinfinite and finite cases.

In the setting of Sec. \ref{WGA-infinite},
%
we have that the step operators $\hat{V}^\dagger$ and $\hat{V}$ satisfy the following unitarity property:
\begin{equation}
 \hat{V}\hat{V}^\dagger=\hat{V}^\dagger\hat{V}=\hat I_{\cal H} \label{unitarity}
\end{equation}

Therefore these operators, together with the number operator $\hat{n}$, constitute
a realization of the Euclidean algebra $E(2)$, with commutators:
\begin{eqnarray}
  \left[\hat{n},\hat{V}\right]  &=& -\hat{V} \nonumber \\
  \left[\hat{n},\hat{V}^\dagger \right] &=& \hat{V}^\dagger \\
  \left[\hat{V},\hat{V}^\dagger \right] &=& 0\nonumber
\end{eqnarray}

It is clear that the unitarity property (\ref{unitarity}) is the eigenvalue equation  for the cuadratic Casimir of the Euclidean algebra $E(2)$, $\hat{C}_2=\hat{V}\hat{V}^\dagger=\hat{V}^\dagger\hat{V}$.

Perelomov-type coherent states \cite{Perelomov86} can be introduced as the action of the Displacement operator:
\begin{equation}
 \hat D(\alpha)=e^{\alpha \hat{V}^\dagger-\alpha^* \hat{V}}\,,\qquad \alpha\in\mathbb{C}\,,
\end{equation}
on the \textit{vacuum} state $|0\rangle$. If $\alpha=re^{i\theta}$, then we have that $\hat D(\alpha)=\hat{U}_\theta(r)$.

The Displacement operator satisfies:
\begin{equation}
  \hat D(\alpha)\hat D(\beta) = \hat D(\alpha+\beta)\,,\qquad \hat D(\alpha)^\dagger=D(-\alpha)
\end{equation}

We define $E(2)$ coherent states as:
\begin{eqnarray}
 |\alpha\rangle&=&\hat D(\alpha)|0\rangle =e^{\alpha \hat{V}^\dagger-\alpha^* \hat{V}}|0\rangle
 = e^{r \left(e^{i\theta}\hat{V}^\dagger- \left( e^{i\theta}\hat{V}^\dagger \right)^{-1}\right)}|0\rangle \nonumber \\
 &=& \sum_{n=-\infty}^{\infty} J_n(2r)   e^{in\theta} \hat{V}^{\dagger n} |0\rangle \\
 &=& \sum_{n=-\infty}^{\infty}
 \alpha^{n}2^{n}k_{n}(2|\alpha|)|n\rangle
 \equiv  \sum_{n=-\infty}^{\infty} c_n(r,\theta) |n\rangle \label{CSE(2)}
\end{eqnarray}
%
where  we have used the generating function of Bessel functions \cite{Gradshteyn}. Here $k_n(x)=\frac{J_n(x)}{x^n}$ are Bochner-Riesz integral kernels \cite{Bochner-Riesz13}.
Thus $E(2)$ coherent states are coherent states of the AN type \cite{GazeauQO19} $|\alpha\rangle =\sum_{n=0}^\infty \alpha^{n} h_n(|\alpha|^2)|n\rangle$ with
$h_n(x)= 2^{n}k_{n}(2 \sqrt{x})$.

Using the asymptotic behaviour of Bessel functions for large $n$:
\begin{equation}
 J_n(x)\approx \frac{1}{\sqrt{2\pi n}} \left(\frac{e x}{2n}\right)^n \label{asymptoticBessel}
\end{equation}
we can check that  $E(2)$ coherent states are defined on the entire complex plane.

The overlap of two coherent states is:
\begin{equation}
 \langle\alpha|\alpha'\rangle = \langle re^{i\theta}|r'e^{i\theta'}\rangle
 = \sum_{n=-\infty}^{\infty} J_n(2r)J_n(2r')e^{in(\theta'-\theta)} = J_0(2 R)
\end{equation}
with $R=\sqrt{r^2+{r'}^2-2rr'\cos(\theta'-\theta)}$, and in the last step we have made use of the \textit{summation theorem} for Bessel functions (see \cite{Gradshteyn}, eq. 8.530).

Particular cases for the overlap are:
\begin{eqnarray}
\langle r e^{i\theta}|r' e^{i\theta} \rangle& = &  J_0(2|r'-r|)  \\
 \langle r e^{i\theta}|r e^{i\theta'} \rangle& = & J_0(4r\sin\left(\frac{\theta'-\theta}{2}\right)  \\
 \langle r e^{i\theta}|r e^{i\theta} \rangle & = & J_0(0)=1
\end{eqnarray}
therefore the states are normalized.

The issue of coherent states of the Eucliean $E(2)$ group has been a  challenge for a long time due to the lack of a resolution of the identity, with many proposals to circumvent it (see \cite{Isham91,DeBievre89} for group theoretical approaches). In \cite{Guerrero21} it was proven that a resolution of the identity can be introduced by means of a nonlocal scalar product originally introduced in \cite{Wolf81} and further developed in \cite{SphereMomentum20}.

The key point is that coherent states of the E(2) group satisfy Helmholtz equation
\begin{equation}
\left( \frac{\partial^2\,}{\partial r^2} + \frac{1}{r} \frac{\partial\,}{\partial r}+
  \frac{1}{r^2} \frac{\partial^2\,}{\partial \theta^2} +k^2\right)|\alpha\rangle=0 \label{HelmholtzPolar}
\end{equation}
with $k=2$, which was derived from the Casimir condition and a differential realization for the E(2) algebra generators \cite{SphereMomentum20}. Coherent states $|\alpha\rangle$ belong to the Hilbert space ${\cal H}_{osc}$ of oscillatory solutions of Helmholtz equaiton \cite{SphereMomentum20}.

Seen the complex number $\alpha=re^{i\theta}=x+i\:y$ as a vector $\vec{\alpha}=(x,y)\in\mathbb{R}^2$, Eq. (\ref{HelmholtzPolar}) is written as
\begin{equation}
 \left[\Delta_{\vec{\alpha}}+k^2 \right]|\alpha\rangle=0\,. \label{HelmholtzCartesian}
\end{equation}

 In Ref. \cite{SphereMomentum20} resolutions of the identity in cartesian and polar coordinates were provided. Here we shall focus on the one in polar coordinates in order to compare with the semiinfinite and finite cases. 

%

 Using the fact that Helmholtz equation is separable  in polar coordinates, and that any regular solution (i.e. without $Y_n$ terms) can be obtained from its values at a circle centered at the origin, we can try to obtain a resolution of the identity involving only coherent states on a circle.
 
 In fact, if $\psi(r,\theta)$ is a solution of (\ref{HelmholtzPolar}) regular at the origin, then
 \begin{equation}
  \psi(r,\theta) = \int_{-\pi}^\pi d\theta' \Delta_{r_0}(\theta-\theta',r)\psic(\theta')
  \label{PropagationHelmholtzPolar}
 \end{equation}
 where $\psic(\theta)=\psi(r_0,\theta)$ and $r_0>0$ is fixed. The Helmholtz propagator in polar
 coordinates is given by:
 \begin{equation}
  \Delta_{r_0}(\theta,r)=\sum_{n\in\mathbb{Z}} \frac{J_n(k r)}{J_n(k r_0)}e^{in\theta}
  \label{PropagatorHelmholtzPolar}
 \end{equation}

 Choosing $0<r_0<\frac{z_{0,1}}{2}$, where $z_{n,j}$ indicates the $j$-th zero of $J_n(x)$ (with the exception of the zero at the origin), it is guaranteed that all Fourier coefficients of $\Delta_{r_0}$ are finite.
 
 Using the asymptotic expression of Bessel functions (\ref{asymptoticBessel}), we have
 that $\frac{J_n(k r)}{J_n(k r_0)}\approx \left(\frac{r}{r_0}\right)^n$ for $n>>1$. Then the Fourier series (\ref{PropagatorHelmholtzPolar}) only converges absolutely when $r<r_0$. For $r=r_0$ the propagator equals a Dirac comb and for  $r>r_0$ eq. (\ref{PropagatorHelmholtzPolar}) should be understood in the context of distributions.
 
 The situation is similar to that of standard coherent states, when any function in phase space is expressed in terms of coherent states on a circle (see \cite{JPAHW12} and references therein). There, the problem is solved restricting the Hilbert space to \textit{physical states}, those with finite values of any moment, since for these states the coefficients in the Fock basis decay fast enough to compensate the divergences appearing in the integration over the circle.
 
 In the present case there is no need to restrict the Hilbert space ${\cal H}_{osc}$, since  regular solutions
 of the Helmholtz equation have Fourier coefficients decaying fast enough to compensate the divergence of the propagator, and thus (\ref{PropagationHelmholtzPolar},\ref{PropagatorHelmholtzPolar}) are well defined for all values of $r$.


Let us introduce \cite{Guerrero21} the following set of states:
\begin{equation}
 |\thetacirc\rangle = |r_0e^{i\theta}\rangle .
\end{equation}
 The overlap of two such states is:

\begin{equation}
 \langle \thetacirc|\thetacirc{}'\rangle  =   \sum_{n=-\infty}^{\infty} J_{n}(2 r_0){}^2e^{in (\theta'-\theta)} = J_0(4r_0\sin\left(\frac{\theta'-\theta}{2}\right) \label{overlapthetaE2}
\end{equation}
%


Then a resolution of the identity in polar coordinates \cite{Guerrero21} is given as:
\begin{equation}
 \hat{A}_{p}= \frac{1}{4\pi^2}\int_{-\pi}^\pi d\theta \int_{-\pi}^\pi d\theta' K(\theta-\theta')
 |\thetacirc{}\rangle\,\langle \thetacirc{}'| = \hat I_{\cal H} \label{ResolutionE2polar}
\end{equation}
where $K(\theta)$ is  the inverse function under convolution of the ovelap (\ref{overlapthetaE2}), given by its Fourier series:
\begin{equation}
 K(\theta)= \sum_{n=-\infty}^{\infty} \frac{1}{J_{n}(2 r_0)^2}e^{in \theta}
\end{equation}

Note that with the choice of $r_0$ as before,  all Fourier coefficients of $K$ are finite. However, they diverge very fast as $n$ increases rendering $K$ a distribution. As with the propagator  (\ref{PropagatorHelmholtzPolar}), taking into account
that $K(\theta)$ will be always integrated multiplied by two regular solutions of Helmholtz equation, the eq. (\ref{ResolutionE2polar}) is well-defined.

The fact that $K(\theta)$ is the inverse under convolution of the overlap (\ref{overlapthetaE2}) guaranties that:
\begin{equation}
\langle \thetacirc| \hat{A}_{p}|\thetacirc{}'\rangle =J_0(4r_0\sin\left(\frac{\theta'-\theta}{2}\right))
\end{equation}
proving that $\hat{A}_{p}$ is in fact the identity since the coherent states in the circle constitute a generating system.


\section{London Coherent States}
\label{Besselstates}



In the setting of Sec. \ref{WGA-semiinfinite}, we have that the step operators $\hat{V}_+,\hat{V}^\dagger_+$ satisfy:

\begin{eqnarray}
 \hat{V}_+\hat{V}^\dagger_+&=&\hat I_{\cal H_+}\\
 \hat{V}^\dagger_+\hat{V}_+&=&\hat I_{\cal H_+}-|0\rangle\langle 0| \label{partialIsometry}
\end{eqnarray}

Therefore these operator are not unitary, but they constitute a partial isometry\footnote{This should be compared with their analogous operators on the {\it extended} Fock basis $\{|n\rangle\,,\,n\in\mathbb{Z} \}$, where they 
are true unitary operators $\hat{V}\hat{V}^\dagger= \hat{V}^\dagger\hat{V}= \hat I_{\cal H}$, and therefore are commuting operators.}. 

These operators are known as  London-Susskind-Glogower \cite{London27,SusskindGlogower64} phase operators, and do not close a finite-dimensional Lie algebra, therefore their treatment will be more complicated that in other cases where the ladder operators close a finite-dimensional Lie algebra, as is the case for the operators $\hat V$ and $\hat V^\dagger$, which close the Euclidean algebra. 

It is clear that the partial isometry property (\ref{partialIsometry}), $\hat{V}_+\hat{V}^\dagger_+=\hat I_{\cal H_+}$, is the eigenvalue equation (with eigenvalue 1) for the quadratic \textit{Casimir}  $\hat{C}^+_2=\hat{V}_+\hat{V}_+^\dagger$ i.e. $\hat{C}_ 2^+=\hat I_{\cal H_+}$. Although there is no finite-dimensional Lie algebra, $\hat{C}^+_2$ commutes with $\hat n$, $\hat V_+$ and $\hat{V}^\dagger_+$, for this reason we still call it a \textit{Casimir}.

We define the displacement operator for London-Susskind-Glogower phase operators $\hat{V}_+,\hat{V}^\dagger_+$ as:
\begin{equation}
 \hat D_{+}(\alpha)=e^{\alpha \hat{V}^\dagger_+-\alpha^* \hat{V}_+}
\end{equation}

This operator can be written as:
\begin{eqnarray}
 \hat D_{+}(\alpha)&=&e^{\alpha \hat{V}^\dagger_+-\alpha^* \hat{V}_+}\nonumber \\
 &=& e^{-\alpha^* \hat{V}_+}e^{\alpha \hat{V}^\dagger_+}
 +\sum_{n=0}^\infty J_{n+2}(2r)\sum_{k=0}^n (-1)^k e^{i(n-2k)\theta}|n-k\rangle\langle k|\\
&=& \sum_{l=0}^\infty\sum_{p=0}^\infty e^{i(l-p)\theta}J_{l-p}(2r)|l\rangle\langle p|  +\sum_{n=0}^\infty J_{n+2}(2r)\sum_{k=0}^n (-1)^k e^{i(n-2k)\theta}|n-k\rangle\langle k|\nonumber\,.
\end{eqnarray}

Although the expression obtained here is slightly different from that of the operator $U_+^\theta(r)$ in Eq. (\ref{Prop-semiinf-twisted}), it can be checked that both expressions coincide.

They satisfy $\hat D_{+}(\alpha)^\dagger =\hat D_{+}(-\alpha)$ and
\begin{equation}
 \hat D_{+}(\alpha)\hat D_{+}(\beta)=\hat D_{+}(\alpha+\beta)
\end{equation}
if $\alpha/\beta\in \mathbb{R}^+_0$, i.e. $\alpha=r e^{i\theta}$ and $\beta=r' e^{i\theta}$. For
different arguments of $\alpha$ and $\beta$ the product does not have a simple expression since the London-Susskind-Glogower phase operators do not close a finite-dimensional Lie algebra.


If $\alpha=re^{i\theta}$, then we have that $\hat D_{+}(\alpha)=\hat{U}^+_\theta(r)$.

We define London coherent states as
\begin{eqnarray} 
 |\alpha\rangle_+&=&\hat D_{+}(\alpha)|0\rangle=e^{\alpha \hat V^\dagger_+-\alpha^*\hat V_+}|0\rangle\nonumber \\
 &=& \sum_{l=0}^\infty e^{i l\theta}J_{l}(2r)|l\rangle + 
 \sum_{n=0}^\infty J_{n+2}(2r) e^{in\theta}|n\rangle\nonumber \\
 &=& \sum_{n=0}^\infty e^{i n\theta}(J_{n}(2r) +  J_{n+2}(2r))|n\rangle \label{BesselCS}\\
 &=& \frac{1}{r}\sum_{n=0}^\infty (n+1)e^{i n\theta}J_{n+1}(2r)|n\rangle  \nonumber\\
 &=& \sum_{n=0}^\infty \alpha^{n}(n+1)2^{n+1}k_{n+1}(2|\alpha|)|n\rangle \equiv \sum_{n=0}^\infty c_n^+(r,\theta) |n\rangle\nonumber
\end{eqnarray}
 Thus London coherent states are coherent states of the AN type \cite{GazeauQO19} $|\alpha\rangle =\sum_{n=0}^\infty \alpha^{n} h_n(|\alpha|^2)|n\rangle$ with
$h_n(x)= (n+1)2^{n+1}k_{n+1}(2 \sqrt{x})$.

As in the case of $E(2)$,  London coherent states are defined on the entire complex plane.


The overlap kernel is given by:
\begin{eqnarray} 
 {}_+\langle r e^{i\theta}|r'e^{i\theta'}\rangle_+=\frac{1}{rr'}\sum_{n=0}^\infty (n+1)^2J_{n+1}(2r)J_{n+1}(2r')e^{in(\theta'-\theta)}
\end{eqnarray}

Using the \textit{summation theorem} for Bessel functions (see \cite{Gradshteyn}, eq. 8.530):
\begin{equation}
 J_0(m R)=\sum_{k=-\infty}^\infty J_k(mr)J_k(mr')e^{ik\varphi}
\end{equation}
where $\varphi=\theta'-\theta$ and $R=\sqrt{r^2+{r'}^2-2rr'\cos\varphi}$, and deriving twice with respect to $\varphi$, we arrite to:
\begin{equation}
  \mathrm{Re}[{}_+\langle r e^{i\theta}|r'e^{i\theta'}\rangle_+]\cos(\theta'-\theta) -  \mathrm{Im}[{}_+\langle re^{i\theta}|r'e^{i\theta'}\rangle_+]\sin(\theta'-\theta)  =
 2 k_1(2R)\cos(\theta'-\theta)  - 8 r r' k_2(2R)\sin^2(\theta'-\theta) \label{ovelap}
\end{equation}

Or, equivalently, 
\begin{equation}
  |{}_+\langle r e^{i\theta}|r'e^{i\theta'}\rangle_+|\cos(\mathrm{Arg}[{}_+\langle re^{i\theta}|r'e^{i\theta'}\rangle_+]+(\theta'-\theta) ) =
 2 k_1(2R)\cos(\theta'-\theta)  - 8 r r' k_2(2R)\sin^2(\theta'-\theta) 
\end{equation}

There are some particular cases where explicit expressions are available:
\begin{eqnarray}
 {}_+\langle r e^{i\theta}|r' e^{i\theta} \rangle_+ & = & 2 k_1(2|r'-r|),\nonumber  \\
\mathrm{Im}[ {}_+\langle r e^{i\theta}|r'e^{i(\theta+\frac{\pi}{2})}\rangle_+] &=&   8 r r' k_2(2\sqrt{r^2+r'{}^2}), \nonumber\\
 {}_+\langle r e^{i\theta}|r' e^{i(\theta+\pi)} \rangle_+ & = & 2 k_1(2(r+r')),  \\
 \mathrm{Im}[{}_+\langle r e^{i\theta}|r'e^{i(\theta+\frac{3\pi}{2})}\rangle_+] &=&  -8 r r' k_2(2\sqrt{r^2+r'{}^2}). \nonumber
\end{eqnarray}

%
Note that for $\theta'=\theta$ or $\theta'=\theta+\pi$ (i.e. $\alpha'/\alpha\in \mathbb{R}$) the scalar product is real.
>From these expressions it is clear that London CS are normalized: ${}_+\langle \alpha|\alpha\rangle_+=1$.


London coherent states constitute an overcomplete family for the Hilbert space $\cal H_+$. 
 In the context of the example of Sec. \ref{WGA-semiinfinite}, any finite energy light distribution in the infinite array, for each value of $z$, can be expanded in terms of the family of coherent states $|\alpha\rangle_L, \alpha\in \mathbb{C}$, apart from the fact that the coherent states
$|z e^{i\theta}\rangle$ represent themselves the  propagation in $z$ of light impinged at the waveguide $n=0$ under the Hamiltonian $\hat H_+^\theta$ given in Eq. (\ref{Ham-semiinf-twisted}).

With the standard construction of a resolution of the identity for Perelomov-type coherent states, we can try with:
%
\begin{eqnarray} 
 \hat A_+&=&\int_\mathbb{C} d^2\alpha\mu(\alpha) |\alpha\rangle_+\langle\alpha|\nonumber\\
&=&\int_0^{\infty}rdr\int_0^{2\pi} d\theta  \mu(re^{i\theta}) |re^{i\theta}\rangle_+ \langle re^{i\theta}|\\
&=&\sum_{n=0}^{\infty}  |n\rangle \langle n|(n+1)^2\int_0^{\infty} dr f(r)\frac{J_{n+1}^2(2r)}{r}
\end{eqnarray}
with some appropriate measure $\mu(\alpha)$, that can be chosen to be of the form $\mu(\alpha)=f(r)$, with 
$r=|\alpha|$	.

We would like to have $\hat A_+=\hat I_{\cal H_+}$, i.e. a Parseval frame like the set of standard coherent states. For this purpose it is required 
that  
\begin{equation}
\int_0^{\infty} dr f(r)\frac{J_{n+1}^2(2r)}{r}=\frac{1}{(n+1)^2}\,,\qquad\forall n\in \mathbb{N}_0
\label{ParsevalCondition}
\end{equation}

This a difficult problem related to the \textit{generalized moment problem}, i.e. to determine a positive weight function $f$ in $[0,\infty)$ such that its generalized moments (with respect to the family $\frac{J_{n+1}^2(2r)}{r}$) are $\frac{1}{(n+1)^2}$.

To circumvent this problem, there are three possibilities: (a) constructing a resolution of the identity with a double integral and a convolution kernel, as in the E(2) case, (b) choosing $f(r)$ such that $\hat A_+$ is at least  invertible with a bounded inverse, (c)  reescaling the coherent states or (d) modifying the London displacement operator
to define modified London coherent states that provide a resolution of the identity.

\subsection{Resolution of the identity in terms of coherent states on the circle}

Let us try the approach (a) mimicking the construction for E(2) coherent states in Sec. \ref{E(2)-CS} (see also \cite{Guerrero21}). A differential realization can also be obtained for the  London-Susskind-Glogower phase operators (although a bit more involved):
\begin{eqnarray}
 \hat{n}_d&=& -i \frac{\partial\,}{\partial\theta}\nonumber\\
 \hat{V}_{+d} &=& -\frac{1}{2}e^{i\theta}\left(\frac{\partial\,}{\partial r} + \frac{i}{r}\frac{\partial\,}{\partial\theta}\right)\frac{\hat n +2}{\hat n +1}= \hat{V}_d \frac{\hat n +2}{\hat n +1} \\
 \hat{V}_{+d}^\dagger &=& \frac{1}{2}e^{-i\theta}\left(\frac{\partial\,}{\partial r} -\frac{1}{r}\left(i\frac{\partial\,}{\partial\theta}-2\right)\right) \frac{\hat n}{\hat n +1} = 
 \left( \hat{V}^\dagger_d + \frac{e^{-i\theta}}{r}\right)\frac{\hat n}{\hat n +1} \nonumber
\end{eqnarray}

Note that the action of these operators on the coefficients of the Coherent States are:
\begin{eqnarray}
 \hat{n}_d\,c_n^+ &=& n c_n^+  \nonumber \\
 \hat{V}_{+d}\, c_n^+ &=& c_{n+1}^+ \\  \nonumber
 \hat{V}^\dagger_{+d}\, c_n^+ &=& c_{n-1}^+\,,\qquad \hat{V}_{+d}^\dagger\, c_0^+=0 \nonumber
\end{eqnarray}
thus the role of lowering and raising operators are interchanged when acting on the coefficients instead of the ket vectors (as it is usual).

It is interesting to note the relation between London coherent states and $E(2)$ coherent states:%
\begin{equation}
 c_n^+= 
 \hat{n}\hat{V}_{+d} \left(\frac{e^{-i\theta}}{r} \hat P_+ c_n\right)
\end{equation}
where $\hat P_+$ is the projector onto the subspace of non-negative $n$ ($\hat P_+ c_n =0$ for $n<0$). 

>From this realization we derive that the eigenvalue equation for the quadratic Casimir, $\hat{C}_ 2^+=\hat{V}_{+d}\hat{V}^\dagger_{+d}=\hat {\cal I}_{{\cal H}_+}$, leads to an heterogeneous  Helmholtz equation:
\begin{equation}
 \hat{C}_ 2^+|\alpha\rangle_{+} =|\alpha\rangle_{+} \qquad \Rightarrow \qquad
%
%
 \left[\vec{\nabla}\cdot (A\vec{\nabla})+k^2 \eta \right]|\alpha\rangle_+=0 \label{HelmholtzLondon}
\end{equation}
where $A(\alpha)=\eta(\alpha)=\alpha^2= r^2 e^{2 i \theta}$, and the wave number is $k=2$. Here $\eta$ plays the role of an heterogeneous index of refraction and $A(\alpha)$ that of a diagonal Riemannian metric.

The heterogeneous character of Eq. (\ref{HelmholtzLondon}) makes very hard to write a resolution of the identity in Cartesian coordinates as in the case of E(2) (see \cite{Guerrero21}), thus we shall restrict for London coherent states to the case of polar coordinates.



Following \cite{SphereMomentum20,Guerrero21}, and using the radial symmetry of this Helmholtz equation, we know that any solution can be determined from its values on a circle centered at the origin. According to this, let us introduce the following set of states:
\begin{equation}
 |\thetacirc\rangle_+ = |r_0e^{i\theta}\rangle_+ 
\end{equation}
with $r_0>0$ to be determined. The overlap of two states is:

\begin{equation}
 {}_+\langle \thetacirc|\thetacirc{}'\rangle_+  =  \frac{1}{r_0^2} \sum_{n=0}^{\infty}(n+1)^2 J_{n+1}(2 r_0){}^2e^{in (\theta'-\theta)} \label{overlapthetaLondon}
\end{equation}
and this can be computed using using eq. (\ref{ovelap}):

\begin{equation}
  \mathrm{Re}[{}_+\langle \thetacirc|\thetacirc{}'\rangle_+]\cos(\theta'-\theta) -  \mathrm{Im}[{}_+\langle \theta|\theta'\rangle\rangle_+]\sin(\theta'-\theta)  =
 2 k_1(4 r_0 \sin\frac{\theta'-\theta}{2})\cos(\theta'-\theta)  - 8 r_0^2 k_2(4 r_0 \sin\frac{\theta'-\theta}{2})\sin^2(\theta'-\theta) 
\end{equation}

Note that the overlap ${}_+\langle \thetacirc|\thetacirc{}'\rangle_+$ is a periodic function of
$\theta'-\theta$ given by a Fourier series (\ref{overlapthetaLondon}).

Then it turns out that a resolution of the identity in polar coordinates is given by
\begin{equation}
\hat{A}_{+p}=\frac{1}{4\pi^2} \int_0^{2\pi}d\theta \int_0^{2\pi}d\theta' K_+(\theta'-\theta)|\thetacirc\rangle_+\langle \thetacirc{}'| = \hat{I}_{\cal H_+}
\label{ResolutionLondonPolar}
\end{equation}
where $K_+(\theta)$ is the inverse function under convolution of the overlap (\ref{overlapthetaLondon}), given by its Fourier series:
\begin{equation}
 K_+(\theta)= r_0^2\sum_{n=0}^{\infty} \frac{1}{(n+1)^2 J_{n+1}(2 r_0)^2}e^{in \theta}
\end{equation}

With the choice of $r_0$ as before, it is guaranteed that all Fourier coefficients of $K_+$ are finite. For the same reason as in the case of Euclidean coherent states, even though the Fourier
coefficients of the kernel $K_+(\theta)$ diverge with $n$, when sandwiched with two  regular
solutions of Helmholtz equation (\ref{HelmholtzLondon}) the result is finite.

\subsection{Non-tight London frame}

Let us try to obtain now a resolution of the identity integrating on the whole complex plane following option (b). For that purpose let us choose $f(r)= \Gamma^2(3/2)/r$ in the integration measure, then 
\begin{eqnarray} 
\hat{A}_+=\Gamma^2(3/2)\sum_{n=0}^{\infty}  |n\rangle \langle n|(n+1)^2\int_0^{\infty} dr \frac{J_{n+1}^2(2r)}{r^2}
\end{eqnarray}

>From Ref. \cite{Gradshteyn}, eq. 6.574:

\begin{eqnarray} 
\int_0^{\infty} dr \frac{J_{n+1}^2(2r)}{r^2}=\frac{\Gamma(n+1/2)}{2\Gamma(n+5/2)}
\end{eqnarray}
because $\Gamma(n+1/2)=\frac{(2n)!}{4^nn!}\sqrt{\pi}$,
\begin{eqnarray}
\int_0^{\infty} dr \frac{J_{n+1}^2(2r)}{r^2}=\frac{4}{\Gamma^2(3/2)(2n+3)(2n+1)}
\end{eqnarray}
so that
\begin{eqnarray}
\hat{A}_+=\sum_{n=0}^{\infty}  |n\rangle \langle n|\frac{4(n+1)^2}{(2n+3)(2n+1)}
\end{eqnarray}
which is a diagonal, invertible operator, with a bounded inverse.

Applying now the general theory of Appendix \ref{frames} to London coherent states defined in Sec. \ref{Besselstates},
we conclude that the set 
$\mathcal{F}_+=\{ |\alpha\rangle_+ \,/\, \alpha\in\mathbb{C}\}$
is a frame with frame operator:
\begin{equation}
 \frac{\Gamma^2(3/2)}{2\pi}\int_\mathbb{C} \frac{d^2\alpha}{|\alpha|} |\alpha\rangle \langle\alpha|
=\hat{A}_+=\sum_{n=0}^{\infty}  |n\rangle \langle n|\frac{4(n+1)^2}{(2n+3)(2n+1)}
\end{equation}

The frame bounds are given by:
\begin{equation}
 \hat I_{\cal H_+}\leq \hat A_+\leq \frac{4}{3} \hat I_{\cal H_+}
\end{equation}
and the frame width is  $w(\mathcal{F}_+)=\frac{1}{7}$. 

In this case, since $A$ is diagonal in the number state basis, the expression of the inverse is
straigthforward:
\begin{eqnarray}
 \hat A^{-1}_+&=&\sum_{n=0}^{\infty}  |n\rangle \langle n|  \frac{(2n+3)(2n+1)}{4(n+1)^2}
 =\sum_{n=0}^{\infty}  |n\rangle \langle n|(1-\frac{1}{4(n+1)^2})\\
 &=& \hat I_{\cal H_+} -\sum_{n=0}^{\infty}  |n\rangle \langle n|\frac{1}{4(n+1)^2}
\end{eqnarray}

The dual frame is
$\tilde{\mathcal{F}}_+=\{ \widetilde{|\alpha\rangle}_+=\hat A^{-1}_+|\alpha\rangle_+ \,/\, \alpha\in\mathbb{C}\}$, with:
\begin{eqnarray}
 \widetilde{|\alpha\rangle}_+&=&\frac{1}{r}\sum_{n=0}^\infty \frac{(2n+3)(2n+1)}{4(n+1)}e^{i n\theta}J_{n+1}(2r)|n\rangle\\
 &=& |\alpha\rangle - \frac{1}{r}\sum_{n=0}^\infty \frac{1}{4(n+1)}e^{i n\theta}J_{n+1}(2r)|n\rangle
\end{eqnarray}

Since $w(\mathcal{F}_+)=\frac{1}{7}$, the series expansion
(\ref{seriesexpansion}) for the reconstruction formula (see Appendix \ref{frames})  converges faster than $\sum_{k=0}^\infty(\frac{1}{7})^k |||\phi\rangle_0||$, and 
$|||\phi\rangle||\leq  |||\phi\rangle_0||$, where
\begin{equation}
 |\phi\rangle_0 =\frac{\Gamma^2(3/2)}{2\pi}\int_\mathbb{C}d^2\alpha  {}_+\langle \alpha|\phi\rangle
 |\alpha \rangle_+
\end{equation}


\subsection{Rescaled London states}

For option (c),
we shall follow \cite{RescaledBasisStates04} and  introduce the diagonal operator:
\begin{equation}
 \hat{R} =\sum_{n=0}^\infty \frac{1}{\sqrt{n+1}}|n\rangle\langle n| 
\end{equation}

Then we introduce the \textit{rescaled London states}: 
\begin{eqnarray}
 |re^{i\theta}\rangle_R &=& \hat{R} |re^{i\theta}\rangle_+ \\
  &=& \sum_{n=0}^\infty\sqrt{n+1}\frac{J_{n+1}(2r)}{r}e^{in\theta}|n\rangle\\
  &=& \sum_{n=0}^\infty \alpha^{n}2^{n+1}\sqrt{n+1}k_{n+1}(2|\alpha|)|n\rangle
\end{eqnarray}

Rescaled London states are coherent states of the AN type \cite{GazeauQO19} $|\alpha\rangle_R =\sum_{n=0}^\infty \alpha^{n} h_n^R(|\alpha|^2)|n\rangle$ with
$h_n^R(x)= 2^{n+1}\sqrt{n+1} k_{n+1}(2 \sqrt{x})$.

Although $\hat{R}$ is bounded and invertible, $\hat{R}^{-1}$ is unbounded. This implies that the family of rescaled London states
will be different to that of London states (\ref{BesselCS}), see \cite{RescaledBasisStates04}.

We can obtain a resolution of the Identity operator using  rescaled London coherent states as
\begin{eqnarray} 
 \int |\alpha\rangle_{R} \langle\alpha|\mu(r)rdrd\theta=\hat{I}_{{\cal H}_+}.
\end{eqnarray}
Integration over $\theta$ yields
\begin{eqnarray} 
 1=(n+1)\int_0^{\infty} \frac{J_{n+1}^2(2r)}{r^2}r\mu(r)dr\,.
\label{eqR1}
\end{eqnarray}
By choosing 
\begin{eqnarray} 
 {\mu(r)}=1
\end{eqnarray}
equation (\ref{eq1}) is reduced to
\begin{eqnarray} 
 1=(n+1)\int_0^{\infty} \frac{J_{n+1}^2(2r)}{r}dr,
\end{eqnarray}
which is true.

Since the operator $\hat{R}$ is not unitary, the rescaled London coherent states are not normalized. This is an undesirable property, for this reason we shall further modify them in order to obtain a suitable family of normalized coherent states verifying a resolution of the identity.

\subsection{Modified London coherent states}

Following option (d) we define modified London states \cite{Curado21} (see also \cite{JMO21} and  \cite{Zelaya21}) as
\begin{eqnarray} \label{coh}
 |\alpha\rangle_{+m}= |r e^{i\theta}\rangle_{+m}=\frac{1}{N(r)}\sum_{n=0}^\infty\sqrt{n+1}J_{n+1}(2r)e^{in\theta}|n\rangle,
\end{eqnarray}
with $N(r)$ a normalization constant:
\begin{eqnarray} \
 {}_{+m}\langle \alpha|\alpha\rangle_{+m}=1=\frac{1}{N(r)^2}\sum_{n=0}^\infty(n+1)J_{n+1}^2(2r).
\end{eqnarray}
It is not difficult to show that
\begin{eqnarray} 
 N^2(r)=2[J_0^2(2r)+J_1^2(2r)]-\frac{1}{r}J_0(2r)J_1(2r).
\end{eqnarray}
Such that we may properly write  London coherent states as
\begin{eqnarray}\label{MLS}
 |\alpha\rangle_{+m} =\frac{1}{\sqrt{2[J_0^2(2r)+J_1^2(2r)]-\frac{1}{r}J_0(2r)J_1(2r)}}\sum_{n=0}^\infty\sqrt{n+1}J_{n+1}(2r)e^{in\theta}|n\rangle,
\end{eqnarray}

Modified London states are coherent states of the AN type \cite{GazeauQO19} $|J(\alpha)\rangle =\sum_{n=0}^\infty \alpha^{n} h_n^J(|\alpha|^2)|n\rangle$ with
$h_n^J(x)= \frac{\sqrt{x}}{N(\sqrt{x})}2^{n+1}\sqrt{n+1} k_{n+1}(\sqrt{x})$.

We can obtain a resolution of the Identity operator using  modified London coherent states as
\begin{eqnarray} 
 \int |\alpha\rangle_{+m} \langle\alpha|\mu(r)rdrd\theta=\hat{I}_{{\cal H}_+}.
\end{eqnarray}
Integration over $\theta$ yields
\begin{eqnarray} 
 1=(n+1)\int_0^{\infty} \frac{r\mu(r)J_{n+1}^2(2r)}{2[J_0^2(2r)+J_1^2(2r)]-\frac{1}{r}J_0(2r)J_1(2r)}dr\,.
\label{eq1}
\end{eqnarray}
By choosing 
\begin{eqnarray} 
 {\mu(r)}=\frac{2}{r^2}[J_0^2(2r)+J_1^2(2r)]-\frac{1}{r^3}J_0(2r)J_1(2r)\,,
\end{eqnarray}
equation (\ref{eq1}) is reduced to
\begin{eqnarray} 
 1=(n+1)\int_0^{\infty} \frac{J_{n+1}^2(2r)}{r}dr,
\end{eqnarray}
which is true.

See \cite{JMO21} for other interesting properties of modified London coherent states.

Note that for the modified London states we can also introduce a unitary  operator $\hat{R}'$ such that 
$|\alpha\rangle_{+m} =\hat{R}' |\alpha\rangle_+$. However, unlike the case of $\hat R$,  $\hat{R}'$ is not a diagonal operator (in the basis 
$|n\rangle$), and, although the general formalism of \cite{RescaledBasisStates04} can still be applied, its properties are more difficult to study. In particular, $\hat{R}'$ involves the operator $\hat{M}|\alpha\rangle = |\alpha||\alpha\rangle$, 
which it is not an invertible operator.

In fact, 
\begin{equation}
 |\alpha\rangle_{+m} =\hat{R}' |\alpha\rangle_+= \frac{\hat{M}}{N(\hat{M})}  \hat{R} |\alpha\rangle_+ 
\end{equation}

\subsection{Relation between modified London coherent states and harmonic oscillator coherent states}
London modifies coherent states (\ref{coh}) may be rewritten as
\begin{eqnarray}
 |\alpha\rangle_{+m} =\frac{J_{\hat{n}+1}(2r)\sqrt{(\hat{n}+1)!}}{\sqrt{N(r)}}\sum_{n=0}^\infty e^{in\theta}\frac{a_+^{\dagger n}}{n!}|0\rangle,
\end{eqnarray}
with $a_+=\sqrt{\hat{n}+1}V_+$. It is easy then to obtain the London modified coherent states by the application of an operator that only depends on the number operator to the a harmonic oscillator coherent state of amplitude one, {\it i.e.},
\begin{eqnarray}
 |\alpha\rangle_{+m} =\frac{e^{\frac{1}{2}}J_{\hat{n}+1}(2r)\sqrt{(\hat{n}+1)!}}{\sqrt{N(r)}}|e^{i\theta}\rangle_{HO}.
\end{eqnarray}

\section{Finite  London Coherent states}

Finally, let us consider the case of a finite number of waveguides. In the setting of Sec. \ref{WGA-finite}, we have that the step operators $\hat{V}_N,\hat{V}^\dagger_N$ satisfy:

\begin{eqnarray}
 \hat{V}_N\hat{V}^\dagger_N&=&\hat I_{\cal H_N}-|N-1\rangle\langle N-1|\\
 \hat{V}_N^\dagger\hat{V}_N&=&\hat I_{\cal H_N}-|0\rangle\langle 0| \label{partialIsometryfinite}
\end{eqnarray}

Therefore these operator are not unitary, and they do not constitute a partial isometry.
They are, however, nilpotent, since  $(\hat{V}_N)^N = (\hat{V}_N^\dagger)^N=\hat{0}$.


Since $\hat{V}_N$ and $\hat{V}^\dagger_N$ do not constitute a partial isometry, unlike the E(2) case and the London case, the Casimir property does not hold, and therefore there is no associated Helmholtz equation. 

As in Secs. \ref{E(2)-CS} and \ref{Besselstates} we can define Perelomov coherent states introducing
 the displacement operator for Finite London  phase operators $\hat{V}_N,\hat{V}_N^\dagger$ as:
\begin{equation}
 \hat D_{N}(\alpha)=e^{\alpha \hat{V}_N^\dagger-\alpha^* \hat{V}_N}
\end{equation}

This operator can be written as:
\begin{eqnarray}
 \hat D_{N}(\alpha)&=&e^{\alpha \hat{V}_N^\dagger-\alpha^* \hat{V}_N} \\
 &=& \sum_{n,m=0}^{N-1} \sum_{l=-\infty}^\infty i^{-(2N+2)l} e^{i(n-m)\theta} \left( J_{n-m-(2N+2)l}(2r)+(-1)^{m} J_{n+m+2-(2N+2)l}(2r) \right)|n\rangle\langle m|\,.\nonumber
\end{eqnarray}

It satisfies $\hat D_{LN}(\alpha)^\dagger =\hat D_{LN}(-\alpha)$ and
\begin{equation}
 \hat D_{N}(\alpha)\hat D_{N}(\beta)=\hat D_{N}(\alpha+\beta)
\end{equation}
if $\alpha/\beta\in \mathbb{R}^+_0$, i.e. $\alpha=r e^{i\theta}$ and $\beta=r' e^{i\theta}$. For
different arguments of $\alpha$ and $\beta$ the product does not have a simple expression since the finite London phase operators do not close a finite-dimensional Lie algebra.


If $\alpha=re^{i\theta}$, then we have that $\hat D_{N}(\alpha)=\hat{U}^\theta_N(r)$.

We define the finite London coherent states as
\begin{eqnarray} 
 |\alpha\rangle_{N}&=&\hat D_{N}(\alpha)|0\rangle=e^{\alpha \hat V_N^\dagger-\alpha^*\hat V_N}|0\rangle\nonumber \\
 &=& \sum_{n=0}^{N-1} e^{i n\theta} \sum_{l=-\infty}^\infty i^{-(2N+2)l}  \left( J_{n-(2N+2)l}(2r)+ J_{n+2-(2N+2)l}(2r) \right)|n\rangle\nonumber \\
 &=& \frac{1}{r} \sum_{n=0}^{N-1} e^{i n\theta}\left( \sum_{l=-\infty}^\infty (-1)^{(N+1)l} (n+1-2(N+1)l)J_{n+1-2(N+1)l}(2r)\right)|n\rangle \nonumber \\
 & =& \sum_{n=0}^{N-1} B^N_n(r) e^{i n\theta}|n\rangle
\end{eqnarray}
with
\begin{equation}
 B^N_n(r)= \frac{1}{r}\sum_{l=-\infty}^\infty (-1)^{(N+1)l} (n+1-2(N+1)l)J_{n+1-2(N+1)l}(2r)\,,\qquad n=0,1,\ldots,N-1
\end{equation}

The functions $ B^N_n(r)$ are well defined as can be checked using eq.  (\ref{asymptoticBessel}), or noting that they can be expressed as a finite sum (from 0 to $N-1$) of products of Chebyshev polynomials (see \cite{SotoEguibar11}). As in the case of E(2)  and London coherent states,  finite London coherent states are defined on the entire complex plane.

The overlap of two coherent states is given by:
\begin{equation}
{}_N\langle re^{i\theta} |r'e^{i\theta'}\rangle_{N} = \sum_{n=0}^{N-1} B^N_n(r)B^N_n(r')e^{i(\theta'- \theta)}\,.
\label{overlapN}
\end{equation}

As in the previous cases, let us define the coherent states on the circle:
\begin{equation}
 |\thetacirc\rangle_N = |r_0e^{i\theta}\rangle_N\,,
\end{equation}
with $r_0>0$ to be determined. The overlap of two states is:

\begin{equation}
 {}_N\langle \thetacirc|\thetacirc{}'\rangle_N  =  \sum_{n=0}^{N-1}B^N_n(r_0)^2 e^{i n (\theta'-\theta)}\,. \label{overlapthetaN}
\end{equation}

Note that the overlap is a trigonometric polynomial in the variable $\theta'-\theta$. Let us fix $N\in\mathbb{N}$  and choose $r_0>0$ such that $B^N_n(r_0)\neq 0$ for $n=0,1,\ldots,N-1$. Define the trigonometric polynomial 
\begin{equation}
 K_N(\theta)=\sum_{n=0}^{N-1}\frac{1}{B^N_n(r_0)^2} e^{-i n \theta}\,.
\end{equation}

Contrary to the cases of $E(2)$ and London coherent states, $K_N(\theta)$ is an ordinary function. Using it we can construct the resolution of the identity
\begin{equation}
 \hat{A}_N=\frac{1}{4\pi^2} \int_{-\pi}^\pi d\theta \int_{-\pi}^\pi d\theta' K_N(\theta-\theta')|\thetacirc\rangle_N\langle \thetacirc{}'| = \hat I_{{\cal H}_N} \,.
\end{equation}

It should be stressed that under the limit $N\rightarrow\infty$ all the construction reproduces that of London coherent states (only the the term $l=0$ survives in the infinite sums). In this limit  $K_N\rightarrow K_+$,
thus $K_N$ can be seen as an approximation of the distribution $K_+$.

The infinite case can also be recovered, with a little more effort, shifting the state $|0>$ to the middle of the finite array, which will be labelled from $-\frac{N-1}{2}$ to $\frac{N-1}{2}$ (in the case of $N$ odd). In the limit $N\rightarrow\infty$ E(2) coherent states are recovered. In this limit  $K_N\rightarrow K$,
thus $K_N$ can also be seen as an approximation of the distribution $K$.

\section*{Acknowledgments}
 JG thanks the support of the Spanish MICINN  through the project PGC2018-097831-B-I00 and Junta de Andaluc\'{\i}a through the project FEDER/UJA-1381026.

\appendix
\section{Appendix: A primer on Frames}
\label{frames}

Let us review the construction of non-tight frames (see for instance \cite{AAG14}).
Consider a set of vectors $\mathcal{F}=\{ |\alpha\rangle \,/\, \alpha\in S\}\subset {\cal H}$, where 
$S=\mathbb{N},\mathbb{Z},\mathbb{R},\mathbb{C},\ldots$, and ${\cal H}$ is a Hilbert space.

Suppose that 
\begin{equation}
 \int_S d\alpha \mu(\alpha)|\alpha\rangle \langle \alpha | = \hat A \,,
\label{resolutionoperator}
\end{equation}
where $\mu(\alpha)$ is an appropriate measure and $\hat{A}$ is an invertible operator with bounded inverse
$\hat A^{-1}$ (this equalities should be understood in the weak sense). This means that 
\begin{equation}
 m ||\phi||^2 \leq \int_S d\alpha\mu(\alpha) |\langle \alpha |\phi\rangle|^2 \leq M ||\phi||^2\,,\qquad 
\forall |\phi\rangle \in {\cal H} \,,
\label{framecondition}
\end{equation}
with $0<m\leq M$.

If $m=M>1$ the frame is \textit{tight}, and in this case $\hat A=m \hat I$. If $m=M=1$ we have a Parseval frame (like the set
of standard coherent states).

The frame condition (\ref{framecondition})  can be written as 
\begin{equation}
 m \hat I\leq \hat A \leq M \hat I
\label{framebounds}
\end{equation}
We shall normally consider $m=inf(Spec(\hat A))$ and $M=sup(Spec(\hat A))$.

\subsection{Dual Frame}

Since $\hat A^{-1}$ is bounded, we can define the new set of vectors 
$\tilde{\mathcal{F}}=\{ \widetilde{|\alpha\rangle} \,/\, \widetilde{|\alpha\rangle}= \hat A^{-1}|\alpha\rangle\,,\alpha\in S\}$.
Then, it can be proven (see, for instance, \cite{AAG14}) that $\tilde{\mathcal{F}}$ is also
a frame with
\begin{equation}
 \int_S d\alpha\mu(\alpha) \widetilde{|\alpha\rangle}\widetilde{ \langle \alpha |} = \hat A^{-1}
\label{resolutionoperatordualframe}
\end{equation}
and frame bounds
\begin{equation}
 M^{-1} ||\phi||^2 \leq \int_S d\alpha\mu(\alpha) |\widetilde{\langle \alpha |}\phi\rangle|^2 \leq m^{-1} ||\phi||^2\,,\qquad 
\forall |\phi\rangle\in {\cal H}
\label{dualframecondition} 
\end{equation}

The most important result is that we can reconstruct the orginal
state $|\phi\rangle$ from its \textit{frame} coefficients $\langle \alpha |\phi\rangle$ using
the dual frame vectors:
\begin{equation}
 |\phi\rangle= \int_S d\alpha\mu(\alpha) \langle \alpha |\phi\rangle \widetilde{|\alpha\rangle} \,,\qquad\forall |\phi\rangle\in {\cal H}
\label{reconstructionformula}
\end{equation}

The frame $\mathcal{F}$ and the dual  frame $\tilde{\mathcal{F}}$ are orthogonal to each
other in the sense that:
\begin{equation}
  \int_S d\alpha \mu(\alpha)|\alpha\rangle \widetilde{ \langle \alpha| } =  
\int_S d\alpha \mu(\alpha)\widetilde{|\alpha\rangle}  \langle \alpha|  = \hat I
\label{frameduality}
\end{equation}

Expression (\ref{frameduality}) is the equivalent to the resolution of the identity
for standard coherent states, where in this case $\hat A=\hat A^{-1}=\hat I$ and thus 
$\widetilde{|\alpha\rangle}=|\alpha\rangle$.

\subsection{Series expansion for the reconstructed states}

Since the reconstruction formula (\ref{reconstructionformula}) involves computing the
 inverse $\hat A^{-1}$ of $\hat A$, it is useful to obtain power series expansions for the inverse
and the reconstruction formulae with control of the size of the different terms of the series.

Let us suppose that the frame $\mathcal{F}$ is \textit{almost} tight, in the sense that $M-m$ is small
and then $m\approx M$. Then $\hat A\approx \frac{m+M}{2} \hat I$, $\hat A^{-1}\approx \frac{2}{m+M} \hat I$ and
therefore $\widetilde{|\alpha\rangle} \approx  \frac{2}{m+M}|\alpha\rangle$. More precisely,
\begin{equation}
 \widetilde{|\alpha\rangle}=\frac{2}{m+M}|\alpha\rangle + (\hat A^{-1}-\frac{2}{m+M}\hat I)|\alpha\rangle
\end{equation}
and plugging this into the reconstruction formula (\ref{reconstructionformula}):

\begin{eqnarray}
 |\phi\rangle&=& \int_S d\alpha\mu(\alpha) \langle \alpha |\phi\rangle (\frac{2}{m+M}|\alpha\rangle 
        +(\hat A^{-1}-\frac{2}{m+M}I)|\alpha\rangle) \nonumber \\
& =& \frac{2}{m+M} \int_S d\alpha\mu(\alpha) \langle \alpha |\phi\rangle|\alpha\rangle
      +(\hat A^{-1}-\frac{2}{m+M}I)\int_S d\alpha\mu(\alpha)  |\alpha\rangle \langle \alpha |\phi\rangle \nonumber \\
 & =& \frac{2}{m+M} \int_S d\alpha\mu(\alpha) \langle \alpha |\phi\rangle|\alpha\rangle
      +(\hat A^{-1}-\frac{2}{m+M}I)A|\phi\rangle \nonumber \\
&=& \frac{2}{m+M} \int_S d\alpha\mu(\alpha) \langle \alpha |\phi\rangle|\alpha\rangle
     + \hat B|\phi\rangle
\end{eqnarray}
where $\hat B=\hat I-\frac{2}{m+M}\hat A$. Solving for $|\phi\rangle$, and denoting 
$|\phi\rangle_0\equiv \hat{A}|\phi\rangle=\int_S d\alpha\mu(\alpha) \langle \alpha |\phi\rangle|\alpha\rangle$, we have:
\begin{equation}
 |\phi\rangle=\frac{2}{m+M} (\hat I-\hat B)^{-1}|\phi\rangle_0 = 
\frac{2}{m+M}\sum_{k=0}^\infty \hat B^k |\phi\rangle_0
\label{seriesexpansion}
\end{equation}

Using (\ref{framebounds}), the operator $\hat B$ verifies:
\begin{equation}
 -\frac{M-m}{m+M}\hat I\leq \hat B \leq \frac{M-m}{m+M}\hat I
\end{equation}
and therefore the operator norm is bounded by $||\hat B||\leq \frac{M-m}{m+M}=w(\mathcal{F})$ where $w(\mathcal{F})$ is
the \textit{frame width}.

The size of each term of the series expansion is bounded by:
\begin{equation}
 ||\hat B^k|\phi\rangle_0|| \leq ||\hat B||^k |||\phi\rangle_0|| \leq w(\mathcal{F})^k |||\phi\rangle_0||
\end{equation}
and therefore its size goes to zero faster than a geometric series of ratio 
$w(\mathcal{F})=\frac{M-m}{m+M}$. Also, 
\begin{equation}
|||\phi\rangle||\leq \frac{2}{m+M} \frac{1}{1-w(\mathcal{F})} |||\phi\rangle_0|| =
\frac{1}{m}|||\phi\rangle_0||
\end{equation}

\bibliographystyle{ieeetr}


\bibliography{../BibliographyCS.bib}

\end{document}